\documentclass[prl,twocolumn,preprintnumbers,superscriptaddress]{revtex4-1}

\usepackage{graphicx}
\usepackage{dcolumn}
\usepackage{amsmath}
\usepackage{color}

\begin{document}

\title{Correlated electronic structure, orbital-selective behavior, and magnetic correlations in double-layer La$_3$Ni$_2$O$_7$ under pressure}

\author{D. A. Shilenko}
\affiliation{Institute of Physics and Technology, Ural Federal University, 620002 Yekaterinburg, Russia}

\author{I. V. Leonov}
\affiliation{M. N. Miheev Institute of Metal Physics, Russian Academy of Sciences, 620108 Yekaterinburg, Russia}
\affiliation{Institute of Physics and Technology, Ural Federal University, 620002 Yekaterinburg, Russia}

\begin{abstract}
Using \emph{ab initio} band structure and DFT+dynamical mean-field theory methods we examine the effects of 
electron-electron interactions on the normal state electronic structure, Fermi surface, and magnetic correlations of the recently discovered double-layer perovskite superconductor La$_3$Ni$_2$O$_7$ under pressure.
Our results suggest the formation of a negative charge transfer mixed-valence state with the Ni valence close to 1.75+. 
We find a remarkable orbital-selective renormalization of the Ni $3d$ bands, with $m^*/m \sim 3$ and 2.3 for the Ni $3z^2-r^2$ and $x^2-y^2$ orbitals, respectively, in agreement with experimental estimates.
Our results for the {\bf k}-dependent spectral functions and Fermi surfaces show significant incoherence of the Ni $3z^2-r^2$ states, implying the proximity of the Ni $3d$ states to orbital-dependent localization. Our analysis of the static magnetic susceptibility suggests the possible formation of the spin and charge (or bond) density wave stripe states. 
\end{abstract}

\maketitle


The recent discovery of superconductivity in the hole-doped infinite-layer nickelate thin films $R$NiO$_2$ ($R$ = La, Nd, Pr, Sr, Ca) \cite{Li_2019,Hepting_2020,Zeng_2020,Osada_2021,Lu_2021,Goodge_2021,Ren_2021,Wang_2021,Pan_2022,Zeng_2022} has stimulated intensive efforts in understanding of the electronic structure, magnetic properties, and microscopic mechanisms of the high-$T_c$ superconductivity in nickelates \cite{Kitatani_2020,Chen_2022a,Nomura_2022,Botana_2022,Gu_2022}. 
Upon various chemical compositions, epitaxial lattice strain, and pressure superconductivity in infinite-layer $R$NiO$_2$ films appears below $T_c \sim 31$ K. In $R$NiO$_2$, Ni ions adopt a nominal 
Ni$^+$ $3d^9$ configuration (with the planar Ni $x^2-y^2$ orbital states dominated near the Fermi level), being isoelectronic to Cu$^{2+}$ in the parent hole-doped superconductor CaCuO$_2$ with a 
critical temperature $\sim$110 K \cite{Azuma_1992,Peng_2017,Savrasov_1996}. However, it has been shown that the low-energy states of $R$NiO$_2$ differ significantly from those of the hole-doped CaCuO$_2$. In fact, in $R$NiO$_2$ the Ni $x^2-y^2$ states experience 
strong hybridization with the rare-earth $5d$ orbitals, resulting in self-doping (with the $5d$ bands crossing the Fermi level) and noncupratelike multi-orbital Fermi surface (FS) \cite{Anisimov_1999,Lee_2004,Botana_2020,Werner_2020,Lechermann_2020a,
Karp_2020a,Karp_2020b,Lechermann_2020b,Wang_2020,Nomura_2020,Si_2020,
Leonov_2020,Ryee_2020,Lechermann_2021,Wan_2021,Leonov_2021,Lechermann_2022,
Malyi_2022,Kutepov_2021,Kreisel_2022}. 

Moreover, recent experiments (resonant 
inelastic X-ray scattering, RIXS) provide evidence for the existence of sizable antiferromagnetic correlations (with no static magnetic order) \cite{Lin_2021,Zhou_2022,Lin_2022} and the formation of translational 
symmetry broken states (charge density wave stripes) \cite{Rossi_2021,Tam_2021,Krieger_2021,Ren_2023,Raji_2023}, implying the importance of strong electronic correlations \cite{Mott_1990,Imada_1998}. Moreover, applications of the DFT+dynamical mean-field theory (DFT+DMFT) \cite{Georges_1996,Kotliar_2006} and GW+DMFT \cite{Sun_2002, Biermann_2003} electronic structure calculations show a remarkable orbital-dependent localization of the Ni $3d$ states, complicated by large hybridization with the rare-earth $5d$ states \cite{Werner_2020,Lechermann_2020a,Karp_2020a,Karp_2020b,Lechermann_2020b,
Wang_2020,Nomura_2020,Si_2020,Leonov_2020,Ryee_2020,Lechermann_2021,Wan_2021,
Leonov_2021,Lechermann_2022,Malyi_2022,Kutepov_2021,Kreisel_2022}.

So far, superconductivity has been observed in the epitaxial thin films of the hole-doped nickelates $R$NiO$_2$ and (undoped) quintuple-layer square-planar nickelate Nd$_6$Ni$_5$O$_{12}$ systems synthesized 
using soft-chemical topotactic reduction of the parent perovskite compounds (with metal hydrides). Interestingly, superconductivity sets in in the thin films nickelates with the Ni ions in a mixed-valence average state near to Ni$^{1.2+}$ \cite{Li_2019,Hepting_2020,Zeng_2020,Osada_2021,Lu_2021,Goodge_2021,Ren_2021,Wang_2021,Pan_2022,Zeng_2022} . However, no evidence of superconductivity has been observed in the bulk samples \cite{Li_2020,Wang_2020_exp,Huo_2022}, while magnetic susceptibility measurements of bulk $R$NiO$_2$ (with $R$ = La, Pr, and Nd) show universal spin-glass behavior at low temperatures \cite{Lin_2022}. This suggests the 
crucial role of hydrogen for superconductivity in the infinite-layer nickelates \cite{Ding_2023}. 

In this respect, the recently reported by Sun \emph{et al.} superconductivity in the (bulk hydrogen free) single crystal double-layer perovskite La$_3$Ni$_2$O$_7$ with a high critical temperature $\sim$80 K under pressure between $\sim$14-43  GPa \cite{Sun_2023,Liu_2023} has stimulated intensive efforts in understanding of its electronic structure \cite{Luo_2023,Lechermann_2023,Christiansson_2023,Yang_2023,
Zhang_2023,Shen_2023b}. In contrast to the infinite-layer nickelates in 
La$_3$Ni$_2$O$_7$ (LNO) the Ni ions adopt a nominal mixed-valence Ni$^{2.5+}$ ($3d^{7.5}$) electron configuration. Under pressure, LNO undergoes a structural transition from the low-pressure $Amam$ to orthorhombic $Fmmm$ phase above $\sim$15 GPa, which is characterized by a change of the bond angle of Ni-O-Ni (to 180$^\circ$ along the $c$ axis, with no tilting of oxygen octahedra) \cite{Liu_2023,Sun_2023}. Superconductivity in the high-pressure $Fmmm$ phase of LNO is found to compete with 
a strange (bad) metal phase above $T_c$ and weakly insulating behavior at pressures below 15 GPa. 

Moreover, in recent experiments (resistance, magnetization, Raman scattering, and specific heat) LNO show anomaly in the transport properties near 153~K (at ambient pressure), suggesting the formation of spin and charge density wave states \cite{Liu_2023}. These experiments also show the importance of effective mass renormalizations caused by correlation effects, implying the significance of electron-electron interactions in LNO. While the electronic state of the high-pressure (HP) phase of LNO
has recently been discussed using various band-structure, GW, DFT+DMFT, and GW+DMFT methods \cite{Luo_2023,Lechermann_2023,Christiansson_2023,Yang_2023,
Zhang_2023,Shen_2023b}, its properties are still poorly understood.
The fundamental question concerning the mechanism of superconductivity and the impact of electronic correlations on superconductivity and magnetism in nickelates remains a subject of intense debates.

We address this topic in our present study. In our work, we explore the effects of correlations on the electronic structure and magnetic state of the normal state of the recently discovered double-layer nickelate superconductor La$_3$Ni$_2$O$_7$ under pressure \cite{Liu_2023,Sun_2023}. In particular, we use the DFT+DMFT approach \cite{Haule_2007,Pourovskii_2007,Leonov_2020b} to study the electronic structure, orbital-dependent correlation effects, Fermi surface topology, and magnetic correlations in LNO. Our results reveal a remarkable orbital-selective renormalizations of the Ni $3d$ states. Our analysis of the {\bf k}-resolved spectral functions and correlated Fermi surfaces suggests significant incoherence of the electronic Ni $3z^2-r^2$ states, implying the proximity of the Ni $e_g$ states to orbital-selective localization. Our results propose the possible formation of spin and charge (or bond) density wave stripe states, which seems to be important for understanding of the anomalous properties of LNO. 

We start by computing the correlated electronic structure and analysis of the orbital-selective behavior of the Ni $3d$ states. In our calculations we use the experimentally refined crystal structure (at about 29.5 GPa) and preform structural optimization of atomic positions using the nonmagnetic DFT \cite{Giannozzi_2009, Giannozzi_2017} (with fixed lattice constants $a$ and $c$). Our DFT results are in overall agreement with those in Refs.~\cite{Luo_2023,Lechermann_2023,Christiansson_2023,Yang_2023,Zhang_2023}. Thus, the partially occupied bands crossing the Fermi level are of Ni $e_g$ character with strong hybridization with the O $2p$ states. The occupied part of the O $2p$ bands appears at -8.5 to -2 eV below the Fermi level. In agreement with previous results, our DFT estimate of the charge-transfer energy difference $\Delta \equiv \epsilon_d - \epsilon_p \sim 3.8$~eV is remarkably smaller than that typical for the infinite-layer nickelates, $\sim$4.2 eV \cite{Li_2019,Hepting_2020,Zeng_2020,Osada_2021,Lu_2021,
Goodge_2021,Ren_2021,Wang_2021,Pan_2022,Zeng_2022,Kitatani_2020,Chen_2022a,Nomura_2022,Botana_2022,Gu_2022}. (It is interesting to note that $\Delta$ estimated taking into account both the occupied and unoccupied parts of the Ni $e_g$ and O $2p$ bands is remarkably smaller, $\sim$1.9 eV). This implies the importance of charge transfer effects in LNO, while the infinite-layer nickelates are in fact more close to a Mott-Hubbard regime \cite{Hepting_2020,Goodge_2021,Kitatani_2020,Chen_2022a,Nomura_2022,Botana_2022,Gu_2022}. 

The bands originating from the La $5d$ states are unoccupied and appear at about 2 eV above $E_F$. We note however that for LNO with the experimental (unrelaxed) crystal structure the La $5d$ states cross the Fermi level near the $\Gamma$-point of the Brillouin zone (BZ). Our DFT+DMFT results for the unrelaxed LNO are summarized in Supplemental Material \cite{Suppl}. This results in self-doping -- charge transfer between the Ni $e_g$ and La $5d$ states similar to that in $R$NiO$_2$. We note that in $R$NiO$_2$ the rare-earth $5d$ states cross the Fermi level near the $\Gamma$ and $A$-points of the BZ, resulting in the appearance of three dimensional FS pockets. Moreover, structural optimization within DFT results in a remarkable upshift of the La $5d$ states
by about 1-2 eV above the $E_F$. As a result, the Fermi surface of the HP LNO (with optimized structure) consists of two 
electron pockets with mixed Ni $x^2-y^2$ and $3z^2-r^2$ character, centered at the $\Gamma$ and $M$ points of the BZ -- a quasi two-dimensional cylinder-like FS at the $\Gamma$ point, with small warping along the $c$ axis, and a more squared-like FS at the $M$ point -- and one hole pocket due to the Ni $3z^2-r^2$ states at the $M$ point.
On the other hand, our DFT calculations for LNO with the experimental structure give an additional quasi-2D cylinder-like electron FS of the La $5d$ character at the $\Gamma$ point, 
with substantial variation of the FS cross section along the $c$ axis (see Supplemental Material Fig.~S1).

\begin{figure}[tbp!]
\centerline{\includegraphics[width=0.5\textwidth,clip=true]{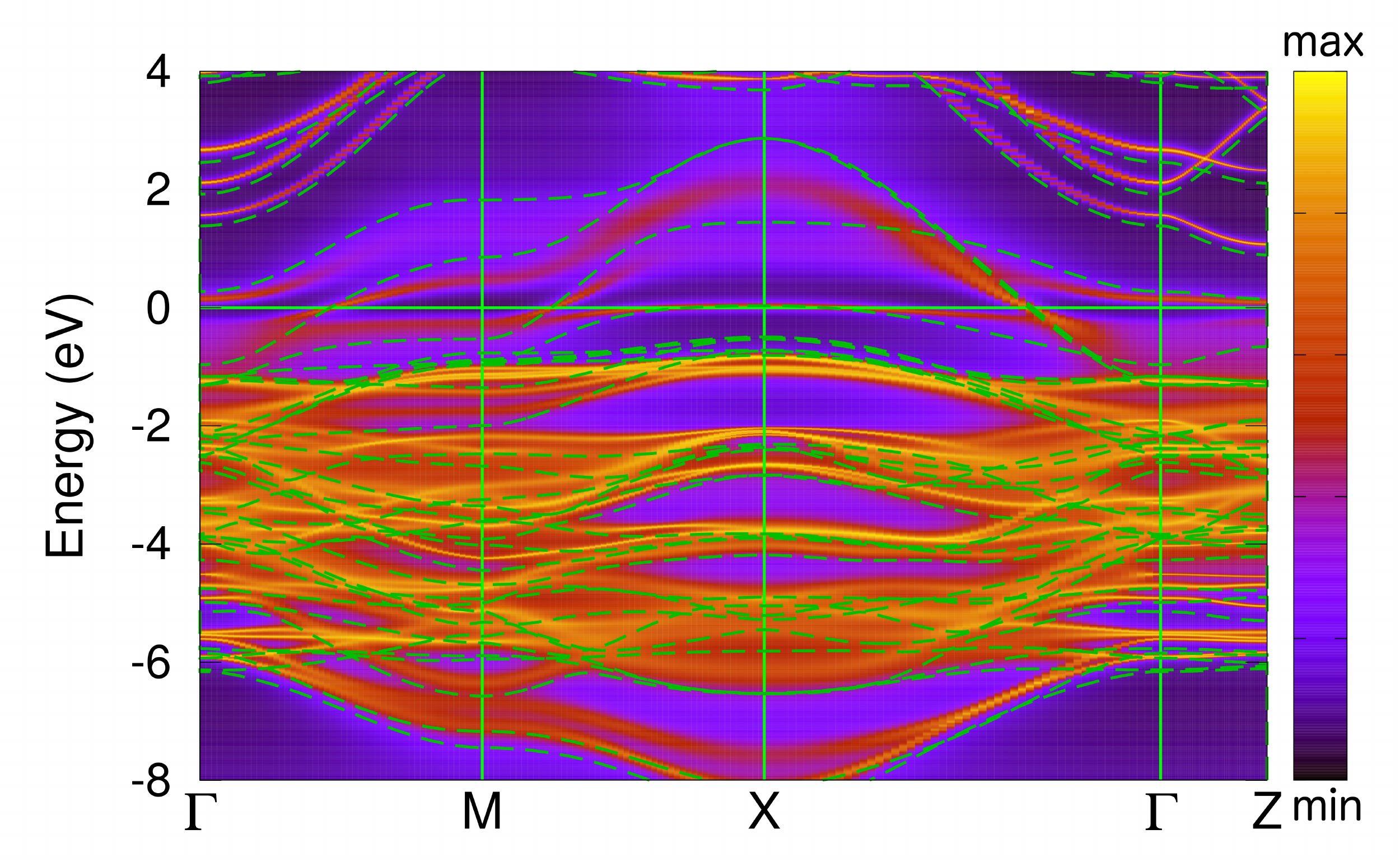}}
\caption{{\bf k}-resolved spectral functions of PM La$_3$Ni$_2$O$_7$ calculated using DFT+DMFT at $T=290$ K. The calculations are performed for the orthorhombic $Fmmm$ structure taken at about 29.5 GPa with optimized atomic positions. The DFT+DMFT spectral functions are compared with the nonmagnetic DFT results (shown with green broken lines).}
\label{Fig_1}
\end{figure}

In order to treat correlation effects in the partially filled Ni $3d$ states (La $5d$ states are more extended, located above $E_F$, and to a large extent can be considered as uncorrelated) we 
employ the DFT+DMFT method \cite{Georges_1996,Kotliar_2006}. We use a fully charge self-consistent DFT+DMFT implementation \cite{Haule_2007,Pourovskii_2007,Leonov_2020b} with plane-wave pseudopotentials \cite{Giannozzi_2009,Giannozzi_2017} to compute the orbital-dependent 
and {\bf k}-resolved spectral functions of paramagnetic (PM) LNO. In these calculations we neglect by the possible appearance of spin and charge density wave states in LNO \cite{Rossi_2021,Tam_2021,Krieger_2021,Ren_2023,Raji_2023,Slobodchikov_2022,
Shen_2023,Chen_2022,Lee_1997,Yoshizawa_2000,Botana_2016,
Zhanga_2016,Bernal_2019,Zhang_2019,Zhang_2020,Hao_2021}. For the low energy states we construct a basis set of atomic-centered Wannier functions for the Ni $3d$, La $5d$, and O $2p$ valence states using the energy window spanned by these bands \cite{Marzari_2012,Anisimov_2005}. This allows us to treat the electron-electron interactions in the partially filled Ni $3d$ shell, complicated by a charge transfer between the Ni $3d$, O $2p$, and La $5d$ states. 
In our DFT+DMFT calculations we employ the continuous-time hybridization expansion (segment) quantum Monte Carlo algorithm \cite{Gull_2011} in order to solve a realistic many-body problem 
describing the strongly correlated Ni $3d$ electrons in LNO. In agreement with previous applications of DFT+DMFT to study infinite-layer and perovskite nickelates \cite{Werner_2020,Lechermann_2020a,Karp_2020a,Karp_2020b,Lechermann_2020b,
Wang_2020,Nomura_2020,Si_2020,Leonov_2020,Ryee_2020,Lechermann_2021,Wan_2021,
Leonov_2021,Lechermann_2022,Malyi_2022,Kutepov_2021,Kreisel_2022} we take the Hubbard $U=6$ eV, Hund's exchange $J = 0.95$ eV, and the fully localized double-counting correction (evaluated from the self-consistently determined local occupations). We neglect the spin-orbit coupling in our calculations. In order to compute the {\bf k}-resolved spectra and correlated Fermi surfaces we perform analytic continuation of the self-energy results using Pad\'e approximants.
Our DFT+DMFT calculations are performed for the normal state of PM LNO at a temperature $T=290$ K.

\begin{figure}[tbp!]
\centerline{\includegraphics[width=0.5\textwidth,clip=true]{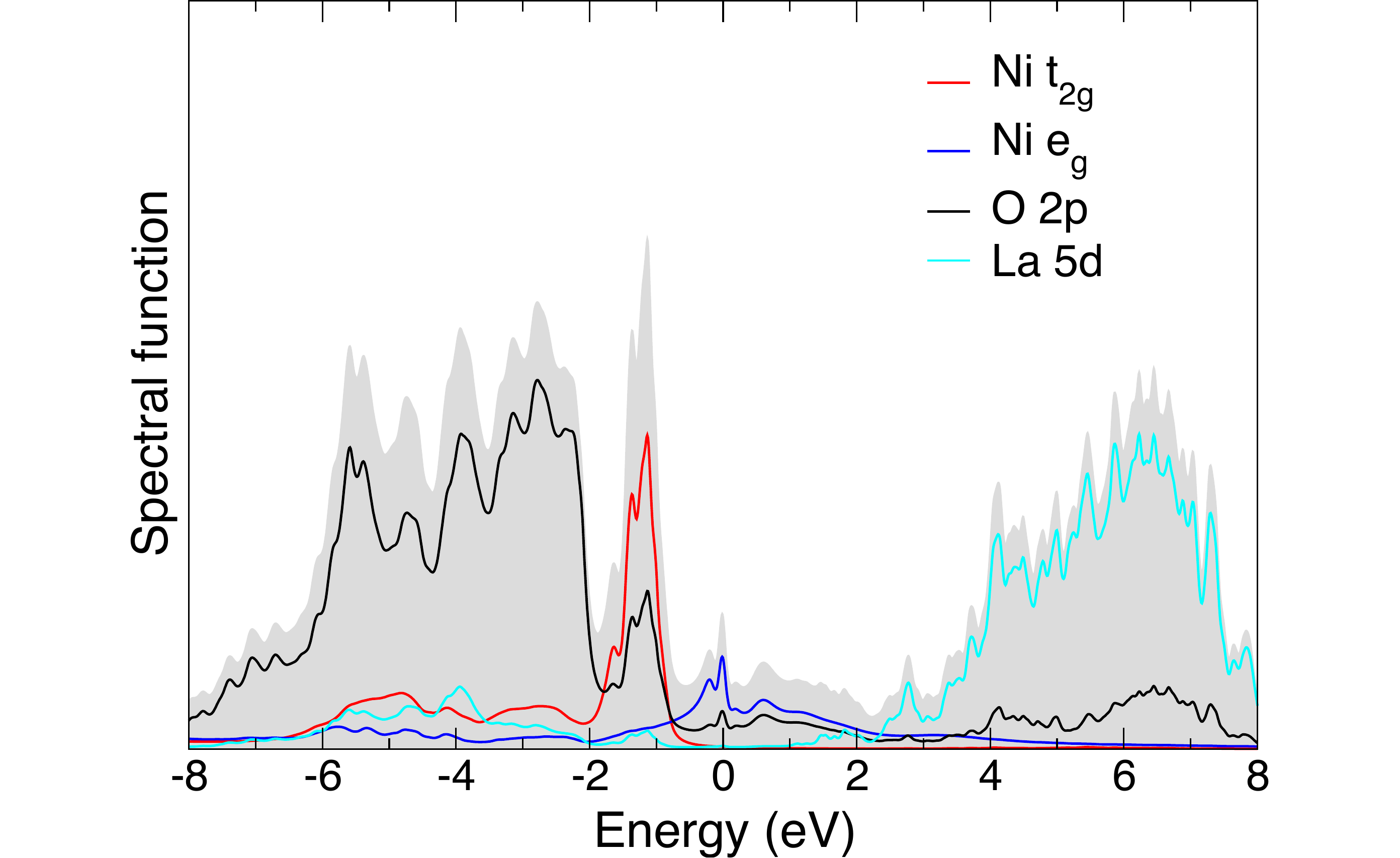}}
\caption{Orbital-dependent spectral functions of PM La$_3$Ni$_2$O$_7$ as obtained by DFT+DMFT at $T=290$ K. }
\label{Fig_2}
\end{figure}

In Fig.~\ref{Fig_1} we display our results for {\bf k}-resolved spectral functions of LNO with the optimized atomic positions obtained using DFT+DMFT (in comparison to the nonmagnetic DFT results). The orbital-dependent spectral functions are shown in Fig.~\ref{Fig_2}. The DFT+DMFT results for LNO with the experimental lattice are shown in SM Figs.~S1 and S2. Overall, our DFT+DMFT results agree well with those published previously \cite{Lechermann_2023,Christiansson_2023}. 
Our calculations show the mixed Ni $x^2-y^2$ and $3z^2-r^2$ states crossing the Fermi level. The occupied O $2p$ states located at -8.5 to -1.9 eV below $E_F$, with strong hybridization with the partially occupied Ni 
$3d$ states. The Wannier Ni $x^2-y^2$ and $3z^2-r^2$ orbital occupations are close to half-filling, of $\sim$0.54 and 0.6 per spin-orbit, respectively. The total Wannier Ni $3d$ occupation is about 8.24. Moreover, 
our analysis of the weights of different atomic configurations of the Ni $3d$ electrons (in DMFT the Ni $3d$ electrons are seen fluctuating between various atomic configurations) gives 0.1, 0.55, and 0.32 for the $d^7$, $d^8$, and $d^9$ configurations, respectively, in accordance with the mixed-valence Ni $3d^8$ and $3d^9$ configurations of the Ni ions. Our analysis gives a formal valence state of 1.75+ for the Ni ions which significantly differ from Ni$^{2.5+}$. Our results therefore suggest that LNO appears close to a negative charge transfer regime \cite{Zaanen_1985,Mizokawa_1991} (similarly with the charge-transfer state in superconducting cuprates), in contrast to the infinite-layer nickelates. The latter are in fact close to a Mott-Hubbard regime \cite{Hepting_2020,Goodge_2021,Kitatani_2020,Chen_2022a,Nomura_2022,Botana_2022,Gu_2022}. Moreover, our estimate for the spin-state configuration gives 0.65 and 0.35 for the high-spin and low-spin state configurations, respectively. The latter suggest strong interplay of the $S=0$ and $S=1/2$ states in the electronic structure of LNO, similarly to that in $R$NiO$_2$.

\begin{figure}[tbp!]
\centerline{\includegraphics[width=0.5\textwidth,clip=true]{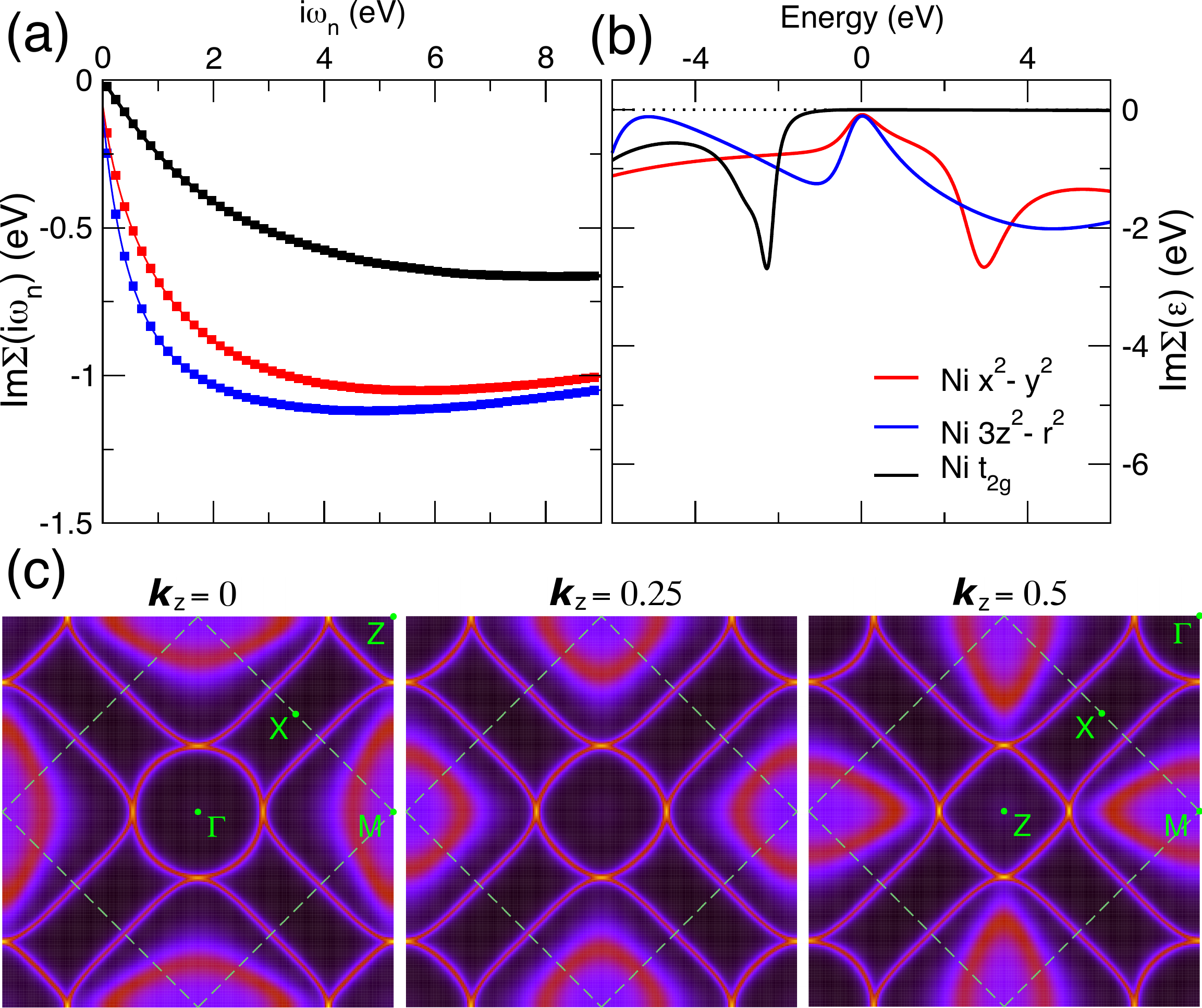}}
\caption{
Orbital-dependent imaginary part of the Ni $3d$ self-energies $\mathrm{Im}[\Sigma(i\omega)]$ on the Matsubara axis (a), 
the imaginary part of the analytically continued Ni $3d$ self-energies $Im[\Sigma(\omega)]$ on the real energy axis (b), and correlated Fermi surfaces [spectral function $A({\bf k},\omega)$ evaluated at $\omega=0$] for different $k_z$ for HP La$_3$Ni$_2$O$_7$ calculated by DFT+DMFT at $T=290$ K.}
\label{Fig_3}
\end{figure}

The electron bands originating from the La $5d$ states are unoccupied and appear at 2.2 eV above $E_F$. Overall, 
the physical picture remains similar to that obtained within DFT, complicated by a remarkable renormalization of the Ni $3d$ states and their substantial orbital-selective incoherence (bad-metal behavior) caused 
by correlation effects. In fact, the electronic structure of LNO is characterized by a Fermi-liquid-like behavior of the Ni $3d$ self-energies [see Fig.~\ref{Fig_3} (a) and (b)], with a substantial orbital-dependent damping of 
$\mathrm{Im}[\Sigma(i\omega)] \sim 0.18$ and 0.25 eV for the Ni $x^2-y^2$ and $3z^2-r^2$ quasiparticle states at the first Matsubara frequency, at $T=290$ K. The Ni $t_{2g}$ states are sufficiently coherent, with $\mathrm{Im}[\Sigma(i\omega)]$ below 0.02 eV at the first Matsubara frequency.  Using Pad\'e extrapolation of the self-energy $\Sigma(i\omega)$ 
to $i\omega \rightarrow 0$ we obtain 0.08 and 0.11 eV for the Ni $x^2-y^2$ and $3z^2-r^2$ states at the Fermi energy, respectively.

\begin{figure}[tbp!]
\centerline{\includegraphics[width=0.5\textwidth,clip=true]{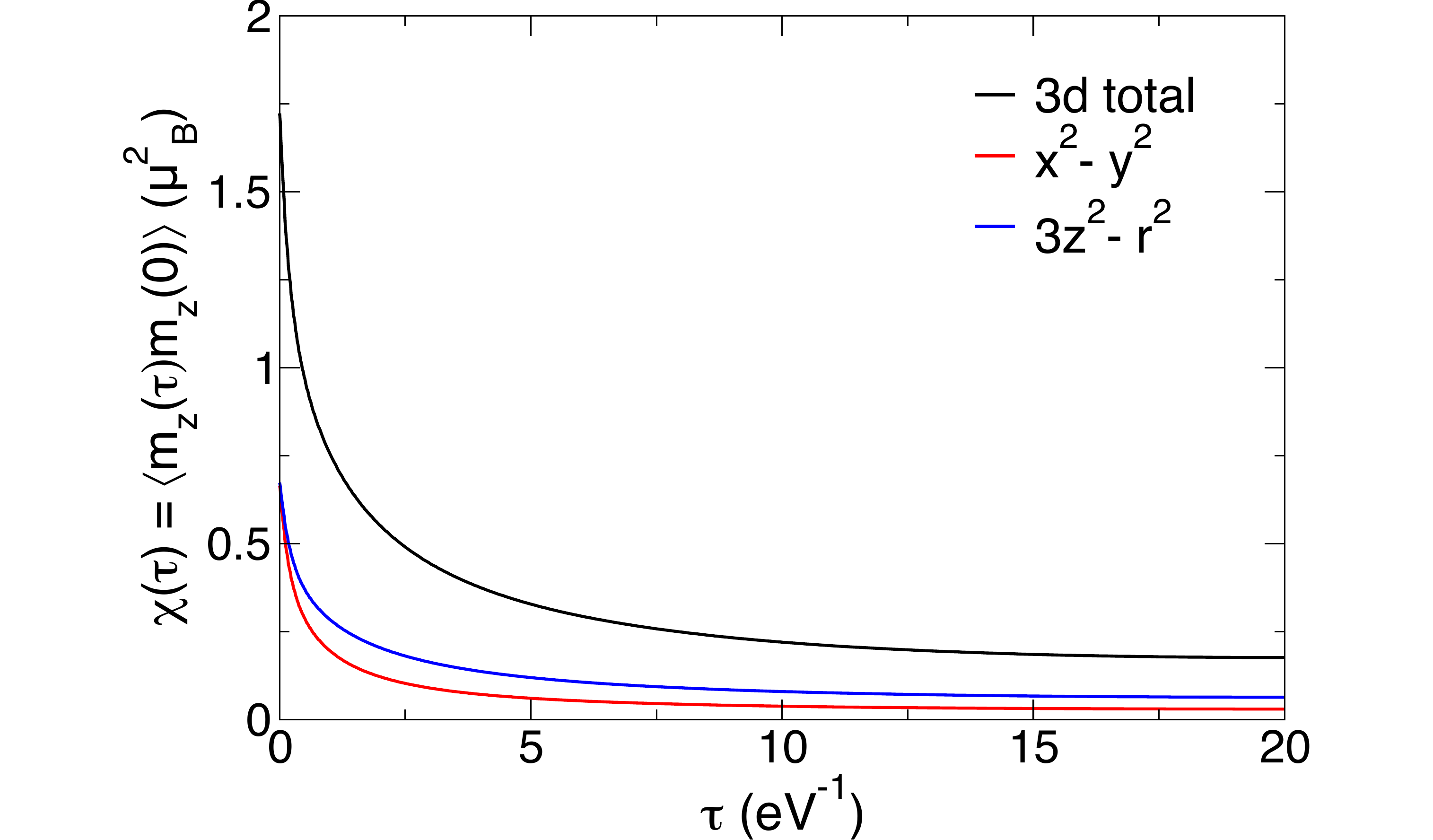}}
\caption{
Orbital-resolved local spin correlation functions $\chi(\tau) = \langle \hat{m}_z(\tau)\hat{m}_z(0) \rangle$ as a function of the imaginary time $\tau$ for the
Ni $3d$ orbitals calculated by DFT+DMFT for the high-pressure PM La$_3$Ni$_2$O$_7$ at $T=290$ K.}
\label{Fig_4}
\end{figure}

Moreover, our DFT+DMFT calculations reveal a remarkable orbital-selective renormalization of the partially occupied Ni $3d$ bands. Our analysis of the orbitally resolved quasiparticle mass enhancement evaluated 
as $\frac{m^*}{m} = 1 - \partial \mathrm{Im}\Sigma(i\omega)/ \partial i\omega |_{\omega\rightarrow 0}$ using Pad\'e approximants gives 2.3 and 3 for the Ni $x^2-y^2$ and $3z^2-r^2$ bands, respectively. The effective mass enhancements of the 
Ni $t_{2g}$ states is much weaker, of $\sim$1.3. That is, the Ni $3z^2-r^2$ states are seen to be more correlated and incoherent-like than the in-plane Ni $x^2-y^2$ orbitals. We note that our result for the quasiparticle 
mass enhancement is in agreement with experimental estimates of $m^*/m$  from transport measurements for single crystals of the double-layer LNO and 
three-layer La$_4$Ni$_3$O$_{10}$, $\sim$2.12 and 2.56, respectively. This behavior is also consistent with our analysis of the orbital-dependent local spin susceptibility 
$\chi(\tau) = \langle \hat{m}_z(\tau) \hat{m}_z(0) \rangle$, evaluated within DMFT (see Fig.~\ref{Fig_4}). Thus, our result suggests the proximity of both Ni $x^2-y^2$ and $3z^2-r^2$ states to localization. While 
the Ni $3z^2-r^2$ orbitals show a slow decaying behavior of $\chi(\tau)$ to 0.06 $\mu_\mathrm{B}^2$ at $\tau=\beta/2$, for the Ni $x^2-y^2$ states it is remarkably smaller, $\sim$0.03 $\mu_\mathrm{B}^2$. Our results therefore suggest 
that magnetic correlations in LNO are at the verge of an orbital-dependent formation of local magnetic moments. In agreement with this, the calculated (instantaneous) magnetic moment of Ni is about 
1.3 $\mu_\mathrm{B}$, which is consistent with a nearly $S = 1/2$ state of nickel. At the same time, the fluctuating moment evaluated as $\mu \equiv [k_\mathrm{B}T \int \chi(\tau) d\tau]^{1/2}$ is significantly smaller, 0.55 $\mu_\mathrm{B}$. It is interesting to mention that our DFT+DMFT calculations for LNO with the experimental structure give similar estimates for $m^*/m$, 2.2 and 2.5, 
for the Ni $x^2-y^2$ and $3z^2-r^2$ orbitals at $T=290$ K, respectively.

\begin{figure}[tbp!]
\centerline{\includegraphics[width=0.5\textwidth,clip=true]{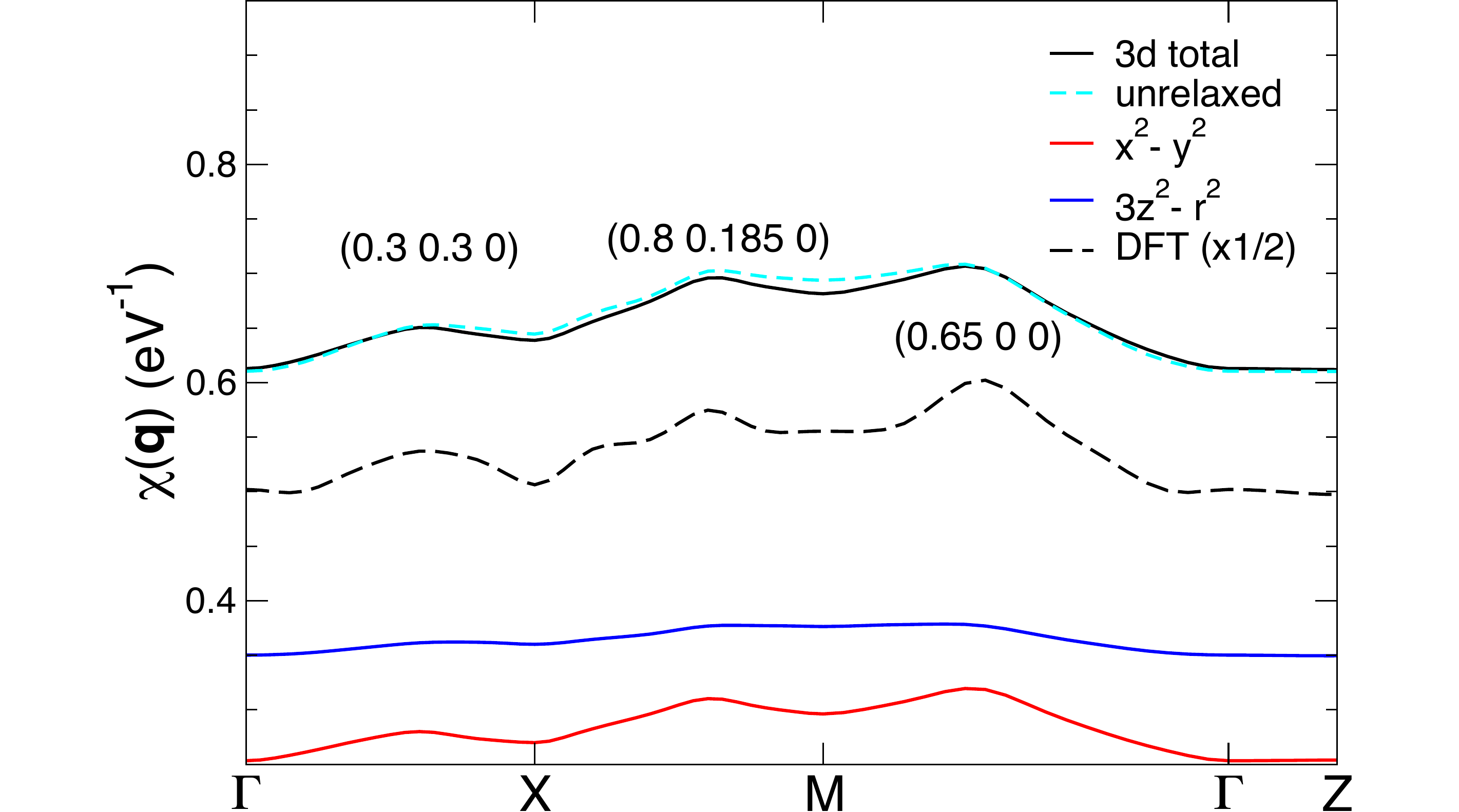}}
\caption{
Orbitally resolved static spin susceptibility $\chi({\bf q})$ of the high-pressure PM La$_3$Ni$_2$O$_7$ obtained by DFT+DMFT at $T=290$ K.}
\label{Fig_5}
\end{figure}

Next, we calculate the correlated Fermi surfaces [spectral functions $A({\bf k},\omega)$ evaluated at $\omega=0$] of PM LNO within DFT+DMFT at $T=290$ K. In Fig.~\ref{Fig_5} we show our results for the in-plane FSs for different $k_z$ (for LNO with the optimized crystal structure). Our results for the FS of LNO with the experimental structure (unrelaxed) are shown in SM Fig. S3.  The DFT+DMFT calculated FSs are similar to that obtained within DFT. In fact, 
it consists of two electron pockets centered at the $\Gamma$ and $M$ points of the BZ and one hole pocket at the $M$ point. However, 
we observe a strong orbital-dependent incoherence of the FS sheets due to correlation effects. Thus, the two electron FS sheets centered at the $\Gamma$ and $M$ points which are of 
mixed Ni $x^2-y^2$ and $3z^2-r^2$ character show more coherent behavior than the hole pocket at the $M$ point, originating from the Ni $3z^2-r^2$ states. We note that this behavior is in overall agreement 
with our analysis of the quasiparticle mass enhancements and local spin susceptibility for the Ni $x^2-y^2$ and $3z^2-r^2$ orbitals. Moreover, for LNO with the experimental 
lattice we obtain nearly similar FSs with an additional (coherent) electron FS sheet, centered at the $\Gamma$ point, of the La $5d$ orbital character. Most interestingly, for $k_z=0.5$ we note a sizeable 
change of the shape of the electron FS sheet centered at the $Z$ point, caused by the lattice effects, from a square- (for the relaxed) to hexagon-like (for the experimental structure). This is suggestive of the 
Pomeranchuk instability \cite{Pomeranchuk_1958,Pomeranchuk_1959,Halboth_2000,Oganesyan_2001,Kitatani_2017}, i.e., a change of the FS shape driven by subtle structural effects. Therefore, this suggests that  structural effects may be important to understand the electronic behavior of LNO. We conclude that correlations effects mainly result in the orbital-dependent effective mass 
renormalizations and incoherence of the spectral weight of LNO.

Our results show multiple in-plane nesting of the calculated Fermi surfaces of PM LNO. We therefore proceed with analysis of the symmetry and strength of magnetic 
correlations in LNO. To this end, we compute the momentum-dependent static magnetic susceptibility $\chi({\bf q})$ in the particle-hole bubble approximation within DFT+DMFT \cite{Skornyakov_2017,Skornyakov_2018}.  
Our results for different orbital contributions in $\chi({\bf q})$ along the BZ $\Gamma$-X-M-$\Gamma$-Z path are shown in Fig.~\ref{Fig_5}. For both the DFT and DFT+DMFT results we observe 
three well-defined maxima of $\chi({\bf q})$ at an incommensurate wave vector $(0.3~0.3~0)$ on the $\Gamma$-X, $(0.8~0.185~0)$  on the X-M, and $(0.65~0~0)$ on the $\Gamma$-M branch of the BZ.
The first anomaly has the lowest value of $\mathrm{max}[\chi({\bf q})]$ and seems to be associated with an electronic spin density wave instability (with a concomitant charge or bond density wave) characterized by 
the $(\frac{1}{3}~\frac{1}{3}~0)$ modulation of the lattice. It is interesting to note that similar behavior was previously discussed 
for the ground state of the hole-doped mixed-valence nickelates (La,Sr)$_2$NiO$_4$ (with Sr $x = 1/3$, Ni$^{2.33+}$ ions) and square-planar La$_4$Ni$_3$O$_8$ with Ni$^{1.33+}$ and La$_3$Ni$_2$O$_6$ with Ni$^{1.5+}$ ions \cite{Lee_1997,Yoshizawa_2000,Botana_2016,Zhanga_2016,
Bernal_2019,Zhang_2019,Zhang_2020,Hao_2021,Lee_1997,Yoshizawa_2000}. Moreover, for hole-doped $R$NiO$_2$ (with a hole-doped Ni$^{+}$ state) our previous DFT+DMFT calculations suggest 
the possible formation of bond-disproportionated striped phase with similar behavior \cite{Slobodchikov_2022,Shen_2023,Chen_2022}. It seems to be consistent with our interpretation of LNO as a negative charge transfer system with the Ni valence state close to 1.75+.

The most pronounced instability of $\chi({\bf q})$ is associated with the $(\frac{2}{3}~0~0)$ modulation, which is strongly competing with a more complex instability with a wave vector $(\frac{13}{16}~\frac{3}{16}~0)$ 
(according to our interpretation). It is important to note that recent experimental studies of infinite-layer nickelates $R$NiO$_2$ also show the evidence for charge density wave state (or lattice modulations) 
with a wave vector $(\frac{1}{3}~0~0)$ \cite{Rossi_2021,Tam_2021,Krieger_2021,Ren_2023,Raji_2023}. It is therefore suggestive for the possible formation of the spin and charge density wave states in LNO, in agreement with recent measurements 
on LNO \cite{Rossi_2021,Tam_2021,Krieger_2021,Ren_2023,Raji_2023,Liu_2023}. This rises the question about the possible role of stripe spin density and charge (or bond) density wave states \cite{Tranquada_1995,Keimer_2015} to explain the electronic properties of LNO. Moreover, we note that our DFT+DMFT calculations predict similar   $\chi({\bf q})$ for LNO with the experimental structure. This implies that this instability is robust and is essentially unaffected by self-doping effects.
Our results suggest complex interplay of spin and charge stripe states, which seems to be important for understanding the anomalous properties of LNO. This topic calls for further theoretical and experimental studies
of the complex interplay between charge order, magnetism, and superconductivity established in nickelate superconductors.


In conclusion, using the DFT+DMFT method we explore the normal state electronic structure, orbital-selective behavior, Fermi surface topology, and magnetic correlations of the recently discover 
double-layer nickelate superconductor La$_3$Ni$_2$O$_7$. Based on our results, we propose that LNO is a negative charge transfer mixed-valence material, with the Ni 
valence state close to 1.75+ (obtained from our analysis of the weights of different atomic configurations of the Ni $3d$ electrons). Our results reveal a remarkable orbital-selective renormalization 
of the Ni $3d$ bands, with $m^*/m \sim 3$ and 2.3 for the Ni $3z^2-r^2$ and $x^2-y^2$ orbitals, respectively, in agreement with experimental estimates. 
Moreover, our results for the {\bf k}-dependent spectral functions and correlated Fermi surfaces show significant incoherence of the electronic Ni $3z^2-r^2$ states. All these imply the proximity of the Ni 
$3d$ states to orbital-dependent localization. Our analysis of the momentum-dependent static magnetic susceptibility suggests the possible formation of the spin and 
charge (or bond) density wave stripe states, which seems to be important for understanding of the anomalous properties of LNO. We propose that superconductivity in double-layer nickelates 
at high pressure is strongly influenced, or even induced, by in-plane spin fluctuations. Our results suggest the emergence of different stripe states on a microscopic level, which interplay
affects the electronic structure and superconductivity of this material.

\begin{acknowledgments}
We acknowledge support by the Russian Science Foundation Project No. 22-22-00926 (https://rscf.ru/project/22-22-00926/).

\end{acknowledgments}


\begin{thebibliography}{100}



\bibitem{Li_2019}
D. Li, K. Lee, B. Y. Wang, M. Osada, S. Crossley, H. R. Lee, Y. Cui, Y. Hikita, and H. Y. Hwang, Nature \textbf{572}, 624 (2019). 

\bibitem{Hepting_2020}
M. Hepting, D. Li, C. Jia, H. Lu, E. Paris, Y. Tseng, X. Feng, M. Osada, E. Been, Y. Hikita \emph{et al.}, Nature Materials \textbf{19}, 381 (2020).

\bibitem{Zeng_2020}
S. Zeng, C. S. Tang, X. Yin, C. Li, M. Li, Z. Huang, J. Hu, W. Liu, G. J. Omar, H. Jani, Z. S. Lim, K. Han, D. Wan, P. Yang, S. J. Pennycook, A. T. S. Wee, and A. Ariando, Phys. Rev. Lett. \textbf{125}, 147003 (2020).

\bibitem{Osada_2021}
M. Osada, B. Y. Wang, B. H. Goodge, S. P. Harvey, K. Lee, D. Li, L. F. Kourkoutis, and H. Y. Hwang, Adv. Mater. \textbf{33}, 2104083 (2021).

\bibitem{Goodge_2021}
B. H. Goodge, D. Li, M. Osada, B. Y. Wang, K. Lee, G. A. Sawatzky, H. Y. Hwang, L. F. Kourkoutis, Proc. Natl. Acad. Sci. U.S.A. \textbf{118}, e2007683118 (2021).

\bibitem{Lu_2021}
H. Lu, M. Rossi, A. Nag, M. Osada, D. F. Li, K. Lee, B. Y. Wang, M. Garcia-Fernandez, S. Agrestini, Z. X. Shen, E. M. Been, B. Moritz, T. P. Devereaux, J. Zaanen, H. Y. Hwang, K.-J. Zhou, W. S. Lee, Science \textbf{373}, 213 (2021).

\bibitem{Wang_2021}
N. N. Wang, M. W. Yang, Z. Yang, K. Y. Chen, H. Zhang, Q. H. Zhang, Z. H. Zhu, Y. Uwatoko, L. Gu, X. L. Dong, K. J. Jin, J. P. Sun, J.-G. Cheng, Nat. Commun. \textbf{13}, 4367 (2022).

\bibitem{Ren_2021}
X. Ren, J. Li, W.-C. Chen, Q. Gao, J. J. Sanchez, J. Hales, H. Luo, F. Rodolakis, J. L. McChesney, T. Xiang \emph{et al.}, 
arXiv:2109.05761


\bibitem{Pan_2022}
G. A. Pan, D. F. Segedin, H. LaBollita, Q. Song, E. M. Nica, B. H. Goodge, A. T. Pierce, S. Doyle, S. Novakov, D. C. Carrizales \emph{et al.}, Nat. Mater. \textbf{21}, 160 (2022).

\bibitem{Zeng_2022}
S. Zeng, C. Li, L. E. Chow, Y. Cao, Z. Zhang \emph{et al.}, Sci. Adv. \textbf{8}, 
abl9927 (2022).

\bibitem{Kitatani_2020}
M. Kitatani, L. Si, O. Janson, R. Arita, Z. Zhong, and K. Held, npj Quantum Materials \textbf{5}, 59 (2020).

\bibitem{Chen_2022a}
H. Chen, A. Hampel, J. Karp, F. Lechermann, and A. J. Millis, Front. Phys. \textbf{10}, 835942 (2022).

\bibitem{Nomura_2022}
Y. Nomura and R. Arita, Rep. Prog. Phys. \textbf{85}, 052501 (2022).

\bibitem{Botana_2022}
A. S. Botana, K.-W. Lee, M. R. Norman, V. Pardo, and W. E. Pickett,
Front. Phys. \textbf{9}, 813532 (2022).

\bibitem{Gu_2022}
Q. Gu and H.-H. Wen, The innovation \textbf{3}, 100202 (2022).



\bibitem{Azuma_1992}
M. Azuma, Z. Hiroi, M. Takano, Y. Bando, and Y. Takeda, Nature (London) \textbf{356}, 775 (1992).

\bibitem{Peng_2017}
Y. Y. Peng, G. Dellea, M. Minola, M. Conni, A. Amorese et al., Nat. Phys. \textbf{13}, 1201 (2017).

\bibitem{Savrasov_1996}
S. Y. Savrasov and O. K. Andersen, Phys. Rev. Lett. \textbf{77}, 4430 (1996).


\bibitem{Anisimov_1999}
V. I. Anisimov, D. Bukhvalov, and T. M. Rice, Phys. Rev. B \textbf{59}, 7901 (1999).

\bibitem{Lee_2004}
K.-W. Lee and W. E. Pickett, Phys. Rev. B \textbf{70}, 165109 (2004).

\bibitem{Botana_2020}
A. S. Botana and M. R. Norman, Phys. Rev. X \textbf{10}, 011024 (2020).


\bibitem{Werner_2020}
P. Werner and S. Hoshino, Phys. Rev. B \textbf{101}, 041104(R) (2020).

\bibitem{Lechermann_2020a}
F. Lechermann, Phys. Rev. X \textbf{10}, 041002 (2020).

\bibitem{Karp_2020a}
J. Karp, A. S. Botana, M. R. Norman, H. Park, M. Zingl, and A. Millis,  Phys. Rev. X \textbf{10}, 021061 (2020).

\bibitem{Karp_2020b}
J. Karp, A. Hampel, M. Zingl, A. S. Botana, H. Park, M. R. Norman, and A. J. Millis, Phys. Rev. B \textbf{102}, 245130 (2020).

\bibitem{Lechermann_2020b}
F. Lechermann, Phys. Rev. B \textbf{101}, 081110(R) (2020).

\bibitem{Wang_2020}
Y. Wang, C.-J. Kang, H. Miao, and G. Kotliar, Phys. Rev. B \textbf{102}, 161118(R) (2020).

\bibitem{Nomura_2020}
Y. Nomura, T. Nomoto, M. Hirayama, and R. Arita, Phys. Rev. Research \textbf{2}, 043144 (2020).

\bibitem{Si_2020}
L. Si, W. Xiao, J. Kaufmann, J. M. Tomczak, Y. Lu, Z. Zhong, and K. Held, Phys. Rev. Lett. \textbf{124}, 166402 (2020).

\bibitem{Ryee_2020}
S. Ryee, H. Yoon, T. J. Kim, M. Y. Jeong, and M. J. Han, Phys. Rev. B \textbf{101}, 064513 (2020).

\bibitem{Leonov_2020}
I. Leonov, S. L. Skornyakov, and S. Y. Savrasov, Phys. Rev. B \textbf{101}, 241108(R) (2020).

\bibitem{Leonov_2021}
I. Leonov, J. Alloys Compd. \textbf{883}, 160888 (2021).

\bibitem{Lechermann_2021}
F. Lechermann, Phys. Rev. Materials \textbf{5}, 044803 (2021).

\bibitem{Wan_2021}
X. Wan, V. Ivanov, G. Resta, I. Leonov, and S. Y. Savrasov, Phys. Rev. B \textbf{103}, 075123 (2021).

\bibitem{Lechermann_2022}
F. Lechermann, Phys. Rev. B \textbf{105}, 155109 (2022).

\bibitem{Malyi_2022}
O. I. Malyi, J. Varignon, and A. Zunger, Phys. Rev. B \textbf{105}, 014106 (2022).

\bibitem{Kutepov_2021}
A. L. Kutepov, Phys. Rev. B \textbf{104}, 085109 (2021).

\bibitem{Kreisel_2022}
A. Kreisel, B. M. Andersen, A. T. R\o{}mer, I. M. Eremin, and F. Lechermann,
Phys. Rev. Lett. \textbf{129}, 077002 (2022).

\bibitem{Lin_2021}
J. Q. Lin, P. V. Arribi, G. Fabbris, A. S. Botana, D. Meyers \emph{et al.}, 
Phys. Rev. Let. \textbf{126}, 087001 (2021).

\bibitem{Zhou_2022}
X. Zhou, X. Zhang, J. Yi, P. Qin, Z. Feng \emph{et al.}, Adv. Mater. \textbf{34}, 2106117 (2022).

\bibitem{Lin_2022}
H. Lin, D. J. Gawryluk, Y. M. Klein, S. Huangfu, E. Pomjakushina, F. von Rohr, and A. Schilling, New J. Phys. \textbf{24}, 013022 (2022).


\bibitem{Rossi_2021}
M. Rossi, M. Osada, J. Choi, S. Agrestini, D. Jost, Y. Lee, H. Lu, B. Y. Wang, K. Lee, A. Nag, Y.-D. Chuang, C.-T. Kuo, S.-J. Lee, B. Moritz, T. P. Devereaux, Z.-X. Shen, J.-S. Lee, K.-J. Zhou, H. Y. Hwang, W.-S. Lee, Nat. Phys. \textbf{18}, 869 (2022).

\bibitem{Tam_2021}
C. C. Tam, J. Choi, X. Ding, S. Agrestini, A. Nag, B. Huang, H. Luo, M. Garc\' ia-Fern\' andez, L. Qiao, K.-J. Zhou, Nat. Mater. \textbf{21}, 1116 (2022).

\bibitem{Krieger_2021}
G. Krieger, L. Martinelli, S. Zeng, L. E. Chow, K. Kummer, R. Arpaia, M. M. Sala, N. B. Brookes, A. Ariando, N. Viart, M. Salluzzo, G. Ghiringhelli, D. Preziosi, Phys. Rev. Lett. \textbf{129}, 027002 (2022).

\bibitem{Ren_2023}
X. Ren, R. Sutarto, Q. Gao, Q. Wang, J. Li, Y. Wang, T. Xiang, J. Hu, F.-Ch. Zhang, J. Chang, R. Comin, X. J. Zhou, and Z. Zhu, arXiv:2303.02865

\bibitem{Raji_2023}
A. Raji, G. Krieger, N. Viart, D. Preziosi, J.-P. Rueff, and A. Gloter, arXiv:2306.10507

\bibitem{Mott_1990} N. F. Mott, \emph{Metal-Insulator Transitions}, Taylor \& Francis, London, 1990.

\bibitem{Imada_1998}
M. Imada, A. Fujimori, Y. Tokura, Rev. Mod. Phys. \textbf{70}, 1039 (1998).


\bibitem{Georges_1996}
A. Georges, G. Kotliar, W. Krauth, and M. J. Rozenberg, Rev. Mod. Phys. \textbf{68}, 13 (1996).

\bibitem{Kotliar_2006}
G. Kotliar, S. Y. Savrasov, K. Haule, V. S. Oudovenko, O. Parcollet, and C. A. Marianetti, Rev. Mod. Phys. \textbf{78},
865 (2006).


\bibitem{Sun_2002}
P. Sun and G. Kotliar, Phys. Rev. B \textbf{66}, 085120 (2002).

\bibitem{Biermann_2003}
S. Biermann, F. Aryasetiawan, and A. Georges, Phys. Rev. Lett. \textbf{90}, 086402 (2003).


\bibitem{Li_2020}
Q. Li, Ch. He, J. Si, X. Zhu, Y. Zhang, and H.-H. Wen, Commun. Mater. \textbf{1}, 16 (2020).

\bibitem{Wang_2020_exp}
B.-X. Wang, H. Zheng, E. Krivyakina, O. Chmaissem, P. P. Lopes, J. W. Lynn, L. C. Gallington, Y. Ren, S. Rosenkranz, J.F. Mitchell, and D. Phelan, Phys. Rev. Materials \textbf{4}, 084409 (2020).

\bibitem{Huo_2022}
M. Huo, Z. Liu, H. Sun, L. Li, H. Lui, C. Huang, F. Liang, B. Shen, and M. Wang, Chin. Phys. B \textbf{31}, 107401 (2022).


\bibitem{Ding_2023}
X. Ding, C. C. Tam, X. Sui, Y. Zhao, M. Xu, J. Choi, H. Leng, J. Zhang, M. Wu, H. Xiao \emph{et al.}
Nature \textbf{615}, 50 (2023).

\bibitem{Sun_2023}
H. Sun, M. Huo, X. Hu, J. Li, Y. Han, L. Tang, Z. Mao, P. Yang, B. Wang, J. Cheng, D.-X. Yao, G.-M. Zhang, and M. Wang, arXiv:2305.09586

\bibitem{Liu_2023}
Z. Liu, H. Sun, M. Huo, X. Ma, Y. Ji, E. Yi, L. Li, H. Liu, J. Yu, Z. Zhang \emph{et al.}
Sci. China: Phys. Mech. Astron. \textbf{66}, 217411 (2023).

\bibitem{Luo_2023}
Z. Luo, X. Hu, M. Wang, W. W\'u, D.-X. Yao, arXiv:2305.15564

\bibitem{Lechermann_2023}
F. Lechermann, J. Gondolf, S. B\"otzel, and I. M. Eremin, arXiv:2306.05121

\bibitem{Christiansson_2023}
V. Christiansson, F. Petocchi, and P. Werner, arXiv:2306.07931

\bibitem{Yang_2023}
Q.-G. Yang, D. Wang, Q.-H. Wang, arXiv:2306.03706

\bibitem{Zhang_2023}
Y. Zhang, L.-F. Lin, A. Moreo, and E. Dagotto, 2306.03231

\bibitem{Shen_2023b}
Y. Shen, M. Qin, and G.-M. Zhang, arXiv:2306.07837


\bibitem{Haule_2007}
K. Haule, Phys. Rev. B \textbf{75}, 155113 (2007).

\bibitem{Pourovskii_2007}
L. V. Pourovskii, B. Amadon, S. Biermann, and A. Georges, Phys. Rev. B \textbf{76}, 235101 (2007).

\bibitem{Leonov_2020b}
I. Leonov, A. O. Shorikov, V. I. Anisimov, and I. A. Abrikosov, Phys. Rev. B \textbf{101}, 245144 (2020).

\bibitem{Giannozzi_2009} 
P. Giannozzi, S. Baroni, N. Bonini, M. Calandra, R. Car \emph{et al.}, J. Phys.: Condens. Matter \textbf{21}, 395502 (2009).

\bibitem{Giannozzi_2017}
P. Giannozzi, O. Andreussi, T. Brumme, O. Bunau, M. B. Nardelli, M. Calandra \emph{et al.}, J. Phys.: Condens. Matter \textbf{29}, 465901 (2017).

\bibitem{Suppl}
See Supplemental Material at http://link.aps.org/supplemental/ for additional figures with our DFT+DMFT results for LNO with the experimental (unrelaxed) crystal structure.


\bibitem{Slobodchikov_2022}
K. G. Slobodchikov and I. V. Leonov, Phys. Rev. B \textbf{106}, 165110 (2022).

\bibitem{Shen_2023}
Y. Shen, M. Qin, and G.-M. Zhang, Phys. Rev. B \textbf{107}, 165103 (2023).

\bibitem{Chen_2022}
H. Chen, Y.-F. Yang, G.-M. Zhang, arXiv:2204.12208.



\bibitem{Lee_1997}
S.-H. Lee and S-W. Cheong, Phys. Rev. Lett. 79, 2514 (1997).

\bibitem{Yoshizawa_2000}
H. Yoshizawa, T. Kakeshita, R. Kajimoto, T. Tanabe, T. Katsufuji, and Y. Tokura, Phys. Rev. B \textbf{61}, 854 (2000).


\bibitem{Botana_2016}
A. S. Botana, V. Pardo, W. E. Pickett, and M. R. Norman,
Phys. Rev. B \textbf{94}, 081105(R) (2016).

\bibitem{Zhanga_2016}
J. Zhang, Y.-S. Chen, D. Phelan, H. Zheng, M. R. Norman, and J. F. Mitchell, Proc. Natl. Acad. Sci. U.S.A. \textbf{113}, 8945 (2016).

\bibitem{Bernal_2019}
O. O. Bernal, D. E. MacLaughlin, G. D. Morris, P.-C. Ho, L. Shu, C. Tan, J. Zhang, Z. Ding, K. Huang, V. V. Poltavets, Phys. Rev. B \textbf{100}, 125142 (2019).

\bibitem{Zhang_2019}
J. Zhang, D. M. Pajerowski, A. S. Botana, H. Zheng, L. Harriger, J. Rodriguez-Rivera, J. P. C. Ruff, N. J. Schreiber, B. Wang, Y.-S. Chen, W. C. Chen, M. R. Norman, S. Rosenkranz, J. F. Mitchell, and D. Phelan, Phys. Rev. Lett. \textbf{122}, 247201 (2019).

\bibitem{Zhang_2020}
J. Zhang, D. Phelan, A. S. Botana, Y.-S. Chen, H. Zheng, M. Krogstad, S. G. Wang, Y. Qiu, J. A. Rodriguez-Rivera, R. Osborn, S. Rosenkranz, M. R. Norman, J. F. Mitchell, Nat. Commun. \textbf{11}, 6003 (2020).

\bibitem{Hao_2021}
J. Hao, X. Fan, Q. Li, X. Zhou, C. He, Y. Dai, B. Xu, X. Zhu, and H.-H. Wen, Phys. Rev. B \textbf{103}, 205120 (2021).


\bibitem{Marzari_2012}
N. Marzari, A. A. Mostofi, J. R. Yates, I. Souza, and D. Vanderbilt, Rev. Mod. Phys. \textbf{84}, 1419 (2012).

\bibitem{Anisimov_2005}
V. I. Anisimov, D. E. Kondakov, A. V. Kozhevnikov, I. A. Nekrasov, Z. V. Pchelkina \emph{et al.}, Phys. Rev. B \textbf{71}, 125119 (2005).


\bibitem{Gull_2011}
E. Gull, A. J. Millis, A. I. Lichtenstein, A. N. Rubtsov, M. Troyer, and P. Werner, Rev. Mod. Phys. \textbf{83}, 349 (2011).

\bibitem{Zaanen_1985}
J. Zaanen, G. A. Sawatzky, and J. W. Allen, Phys. Rev. Lett. \textbf{55}, 418 (1985).

\bibitem{Mizokawa_1991}
T. Mizokawa, H. Namatame, A. Fujimori, K. Akeyama, H. Kondoh, H. Kuroda, and N. Kosugi,
Phys. Rev. Lett. \textbf{67}, 1638 (1991).

\bibitem{Pomeranchuk_1958}
I. Ya. Pomeranchuk,  J. Exptl. Theoret. Phys. (U.S.S.R.) \textbf{3}, 524 (1958).

\bibitem{Pomeranchuk_1959}
I. Ya. Pomeranchuk, Sov. Phys. JETP \textbf{8}, 361 (1959).

\bibitem{Halboth_2000}
Ch. J. Halboth and W. Metzner, Phys. Rev. Lett. \textbf{85}, 5162 (2000).

\bibitem{Oganesyan_2001}
V. Oganesyan, S. A. Kivelson, and E. Fradkin, Phys. Rev. B \textbf{64}, 195109 (2001).

\bibitem{Kitatani_2017}
M. Kitatani, N. Tsuji, and H. Aoki, Phys. Rev. B \textbf{95}, 075109 (2017).


\bibitem{Skornyakov_2017}
S. L. Skornyakov, V. I. Anisimov, D. Vollhardt, and I. Leonov
Phys. Rev. B \textbf{96}, 035137 (2017).

\bibitem{Skornyakov_2018}
S. L. Skornyakov, V. I. Anisimov, D. Vollhardt, and I. Leonov
Phys. Rev. B \textbf{97}, 115165 (2018).


\bibitem{Tranquada_1995}
J. M. Tranquada, B. J. Sternlieb, J. D. Axe, Y. Nakamura, and S. Uchida, Nature \textbf{375}, 561 (1995).

\bibitem{Keimer_2015}
B. Keimer, S. A. Kivelson, M. R. Norman, S. Uchida, and J. Zaanen, Nature \textbf{518}, 179 (2015).



\end{thebibliography}
\end{document}